\documentclass{IEEEtaes}

\usepackage{color,array,amsthm}
\usepackage{graphicx}
\usepackage{orcidlink}
\usepackage{amsmath}
\usepackage{amssymb}
\usepackage{algorithm}
\usepackage{algpseudocode}
\usepackage{cite}
\usepackage{subfig}
\usepackage{main}
\usepackage{nicefrac}
\usepackage{tikz}

\usetikzlibrary{positioning,arrows.meta,calc}
\usepackage{pgfplots}
\pgfplotsset{compat=1.18}
\usepgfplotslibrary{polar}

\usetikzlibrary{external}
\pgfplotsset{
    every axis plot/.append style={line join=round}
}% default is line join=miter.

\jvol{XX}
\jnum{XX}
\jmonth{XXXXX}
\paper{1234567}
\pubyear{2026}
\doiinfo{TAES.2026.Doi Number}

\newcommand{\mailto}[1]{\href{mailto:#1}{#1}}
\graphicspath{{../Code/Figures/}}
\begin{document}

\title{Bernoulli Filtering for Multi-Sensor Tracking with Thresholded Measurements}

\author{GUSTAV ZETTERQVIST \orcidlink{0000-0001-6672-4472}} 
% \member{Fellow, IEEE}
% \affil{Department of Electrical Engineering, Linköping University, Linköping, Sweden} 

\author{FREDRIK GUSTAFSSON \orcidlink{0000-0003-3270-171X}}
\member{Fellow, IEEE}
% \affil{Department of Electrical Engineering, Linköping University, Linköping, Sweden}

\author{GUSTAF HENDEBY \orcidlink{0000-0002-1971-4295}}
\member{Senior Member, IEEE}
\affil{Department of Electrical Engineering, Linköping University, Linköping, Sweden} 

%% \author{FOURTH D. AUTHOR}
%% \affil{University of Colorado, Colorado, USA}

% This paragraph of the first footnote will contain the date on which you submitted your paper for review, which is populated by IEEE. It is IEEE style to display support information, including sponsor and financial support acknowledgment, here and not in an acknowledgment section at the end of the article. For example, ``This work was supported in part by the U.S. Department of Commerce under Grant BS123456.'' }
\receiveddate{Manuscript received XXXXX 00, 0000; revised XXXXX 00, 0000; accepted XXXXX 00, 0000.\\
This work was supported in part by the Wallenberg AI, Autonomous Systems and Software Program (WASP) through the Knut and Alice Wallenberg Foundation; and in part by Excellence Center at Linköping-Lund in Information Technology (ELLIIT).}
%% \accepteddate{XXXXX XX XXXX}
%% \publisheddate{XXXXX XX XXXX}

% \corresp{The name of the corresponding author appears after the financial information, e.g. {\itshape (Corresponding author: M. Smith)}. Here you may also indicate if authors contributed equally or if there are co-first authors.}

\authoraddress{
Authors' addresses:
The authors are with the Department of Electrical Engineering, Linköping University, 581 83 Linköping, Sweden,
E-mail: (\mailto{gustav.zetterqvist@liu.se}, \mailto{fredrik.gustafsson@liu.se}, \mailto{gustaf.hendeby@liu.se}).
\textit{(Corresponding author: Gustav Zetterqvist)}.}

% The next few paragraphs should contain the authors' current affiliations, including current address and e-mail. For example, First A. Author is with the National Institute of Standards and Technology, Boulder, CO 80305 USA 
% (e-mail: \href{mailto:author@boulder.nist.gov}{author@boulder.nist.gov}). Second B. Author, Jr., was with Rice University, Houston, TX 77005 USA. He is now with the Department of Physics, Colorado State University, Fort Collins, CO 80523 USA (e-mail: \href{mailto:author@lamar.colostate.edu}{author@lamar.colostate.edu}). Third C. Author is with the Electrical Engineering Department, University of Colorado, Boulder, CO 80309 USA, on leave from the National Research Institute for Metals, Tsukuba 305-0047, Japan 
% (e-mail: \href{mailto:author@nrim.go.jp}{author@nrim.go.jp}).}

% \editor{Mentions of supplemental materials and animal/human rights statements can be included here.}
% \supplementary{Color versions of one or more of the figures in this article are available online at \href{http://ieeexplore.ieee.org}{http://ieeexplore.ieee.org}.}

\markboth{ZETTERQVIST ET AL.}{Bernoulli Filtering for Multi-Sensor Tracking with Threshold-Induced Detection Probability}
\maketitle
\glsdisablehyper

\begin{abstract}
Target tracking is challenging when sensor detection thresholds cause state-dependent missed detections, particularly in multi-sensor scenarios with clutter and uncertain target existence. A recently developed missed detection framework models detection probability as a function of target state, sensor characteristics, and detection threshold, but it is limited to individual measurements and does not address the recursive tracking problem. This work extends the framework using a \gls{bf} formulation to jointly handle recursive target tracking, clutter, and target existence uncertainty. A \gls{pf} implementation of the \gls{bf} is evaluated in a simulated 2D multi-sensor tracking scenario with nonlinear measurements, clutter, and detection uncertainty. Incorporating accurate detection threshold knowledge reduces the \gls{gospa} metric by 53.9\% compared to a conventional \gls{bf} with fixed detection probability, and 35.7\% compared to a \gls{bf} with unknown detection probability, while better balancing missed detections and false alarms.
\end{abstract}

\begin{IEEEkeywords}
Bernoulli filter, particle filter, target tracking, missed detection, false alarms, detection threshold, random finite sets, multi-sensor fusion, GOSPA metric, Received signal strength, RSS measurements
  % Enter keywords or phrases in alphabetical order, separated by commas. For a list of suggested keywords, send a blank e-mail to \href{mailto:keywords@ieee.org}{keywords@ieee.org} or visit \href{http://www.ieee.org/organizations/pubs/ani\_prod/keywrd98.txt}{\url{http://www.ieee.org/organizations/pubs/ani\_prod/keywrd98.txt}}
\end{IEEEkeywords}

\glsresetall
% \tableofcontents

\section{INTRODUCTION}\label{sec:intro}
State estimation in the presence of missed detections and false alarms is a fundamental problem in target tracking. The \gls{bf} has emerged as a powerful framework for single-target tracking in scenarios where the target may or may not exist and where measurements are affected by clutter and missed detections. The underlying Bernoulli \gls{rfs} formulation models uncertainty in both target existence and target state \cite{mahler2007statistical}, and practical Bernoulli filtering recursions for tracking in cluttered environments were later derived in \cite{ristic2013tutorial}. Since then, the \gls{bf} has become an important component in \gls{rfs} based target tracking frameworks \cite{vo2012multi,ristic2012bernoulli,kim2021bernoulli}.

Several extensions of the Bernoulli and related multi-Bernoulli filtering frameworks have been proposed for practical tracking problems. Track-before-detect formulations based on the \gls{bf} have been developed for low \gls{snr} detection and tracking using sensor measurements \cite{papi2014bernoulli}, and labeled multi-Bernoulli track-before-detect particle filters have been proposed for multi-target scenarios \cite{garcia2016lmbtbd}. Bernoulli filtering has also been used for joint detection and direction-of-arrival tracking problems in cluttered environments \cite{zhang2018joint}. Multi-sensor extensions have been developed for fusion of measurements from multiple sensors \cite{saucan2017multisensor}. In addition, unknown detection probability has been addressed through robust \gls{pmbm} filtering formulations that estimate the detection probability online \cite{li2021robustpmbm}, and probabilistic Bernoulli filters have been proposed to account for imprecise detection models and sensor uncertainty \cite{houssineau2020possibilistic}. However, these approaches typically do not explicitly account for sensor detection thresholds and their induced state-dependent probability of detection.

In our previous work \cite{zetterqvist2025misseddetection,zetterqvist2025DirectionalSensitivity}, a missed detection framework that models the probability of detection as a function of the target state and sensor characteristics, including the detection threshold, is proposed. This framework provides insight into the impact of detection thresholds on tracking performance, but it does not account for false alarms or target existence uncertainty, which limits its applicability in realistic cluttered environments.

In this work, that framework is extended by formulating the problem within the \gls{bf} framework. This allows false alarms to be modeled through a \gls{ppp} while also accounting for target existence uncertainty through a random variable. The resulting recursive filter updates both the target existence probability and the posterior spatial density of the target state using measurement sets from multiple sensors.

While previous Bernoulli- and \gls{rfs} based filters have considered uncertain or unknown detection probabilities, no prior work explicitly incorporates the effect of sensor detection thresholds on the probability of detection within the \gls{bf} recursion. This work fills this gap by deriving the \gls{bf} equations with a measurement model that includes threshold-dependent detection probability, and evaluating the resulting \gls{pf} implementation of the \gls{bf} in a simulated multi-sensor tracking scenario with nonlinear measurements, clutter, and state-dependent detection probability. The results show that explicitly modeling the detection threshold improves tracking performance by better balancing the trade-off between missed detections and false alarms.

The paper is organized as follows. 
The problem formulation is given in \Secref{sec:prob_form}. Relevant \gls{rfs} models, including the \gls{ppp} clutter model and the Bernoulli \gls{rfs}, are reviewed in \Secref{sec:background}. The measurement model with state- and threshold-dependent detection probability is defined in \Secref{sec:method}, followed by the derivation of the recursive Bernoulli filtering equations in \Secref{sec:filter_eqs}. The simulation setup is described in \Secref{sec:sim_model}, and the corresponding results are presented and discussed in \Secref{sec:sim_results}. Finally, concluding remarks are given in \Secref{sec:conclusion}.

\section{Problem Formulation}\label{sec:prob_form}

In this work, the problem of single-target tracking in a multi-sensor environment is considered. Each sensor has a detection threshold that affects the probability of detection.
The target state at time $k$ is denoted by $x_k \in \mathbb{R}^{n_x}$ and evolves according to the dynamic model
$$ x_k = f(x_{k-1}) + w_k, $$
where $f(\cdot)$ is a (possibly nonlinear) state transition function and $w_k$ is process noise with known distribution.

A set of $S$ sensors provides measurements of the target. The measurement from sensor $s$ at time step $k$ is modeled as
$$z_k^{(s)} = h^{(s)}(x_k) + e_k^{(s)}, $$
where $h^{(s)}(\cdot)$ denotes the measurement function associated with sensor $s$, and $e_k^{(s)}$ is zero-mean measurement noise with a known distribution. 
In practice, each sensor produces a finite set of measurements $Z_k^{(s)}$ at every time step, which may contain at most one target-generated measurement together with an unknown number of clutter measurements.

The probability of detection, \ie, the probability that the measurement set $Z_k^{(s)}$ contains a target-generated measurement, is modeled as a function of the target state and a detection threshold $\gamma$, denoted by~$P_D(x_k; \gamma)$.
% Although the formulation allows for sensor-specific thresholds, \ie, $\gamma^{(s)}$, a common threshold is used for all sensors in this work.

Measurements not originating from the target are modeled as clutter, represented by a \gls{ppp} with a known intensity function $\kappa(z)$ for each sensor. 
Furthermore, the target is assumed to exist only intermittently and may be absent at any given time.

The problem is formulated within the \gls{bf} framework, where the target is modeled as a Bernoulli \gls{rfs} characterized by an existence probability~$q_k$ and a spatial density~$p_k(x)$.

The objective is to recursively estimate the posterior of~$(q_k, p_k(x))$ given the sequence of multi-sensor measurements ${\mathcal{Z}_1, \dots, \mathcal{Z}_k}$, where $\mathcal{Z}_k = \big( Z_k^{(1)}, \ldots, Z_k^{(S)} \big)$, while explicitly incorporating the threshold-dependent detection probability into the measurement likelihood.

Key notation used throughout the paper is summarized in \Tabref{tab:notation} for reference.

\begin{table}[tb]
    \centering
    \caption{Summary of key notation used in the paper.}\label{tab:notation}
    \begin{tabular}{lp{0.7\columnwidth}} 
        \toprule
        \textbf{Notation} & \textbf{Description} \\
        \midrule
    $X_k$ & Bernoulli \gls{rfs} representing the target state at time~$k$ \\
    $Z_k$ & Finite set of measurements at time $k$ \\
    $\pi(X)$ & Probability density function of a \gls{rfs} $X$ \\
    $\phi(X_k \mid X_{k-1})$ & Transition density of a \gls{rfs} from time~$k-1$ to~$k$ \\
    $g(Z_k \mid X_k)$ & Measurement likelihood of a \gls{rfs}~$Z_k$ given the state~$X_k$ \\
    $x_k$ & State of the target at time~$k$ \\
    $p(x)$ & Probability density function of a single-target state~$x$ \\
    $f(x_k \mid x_{k-1})$ & State transition density of a single target from time~$k-1$ to~$k$ \\
    $\ell(z \mid x, \gamma)$ & Measurement likelihood of a single measurement~$z$ given the target state~$x$ and threshold~$\gamma$ \\
    $q_{k|k}$ & Existence probability of the target at time~$k$ \\
    $p_{k|k}(x)$ & Spatial density of the target state at time~$k$ \\
    $P_D(x;\gamma)$ & Probability of detection given state~$x$ and threshold~$\gamma$ \\
    $\lambda$ & Clutter intensity (expected number of false alarms) \\
    $c(z)$ & Clutter spatial distribution \\
    $\kappa(z)$ & Clutter intensity function, defined as $\kappa(z) \! = \!\lambda c(z)$ \\
    $p_b$ & Birth probability of the target \\
    $p_s$ & Survival probability of the target \\
    \bottomrule
    \end{tabular}
\end{table}
\section{Background}\label{sec:background}
This section provides an overview of relevant \gls{rfs} models used in our formulation, including the \gls{ppp} for clutter modeling and the Bernoulli \gls{rfs} for single-target tracking.

\subsection{Random Finite Set Models}
The \gls{rfs} framework provides a probabilistic formulation of problems involving an unknown number of objects and measurements. An \gls{rfs} is a set-valued random variable with random cardinality and random elements~\cite{mahler2007statistical}. This makes it suitable for modeling both target states and measurement sets in tracking problems.

In this work, we consider the single-target case and introduce the \gls{rfs} models required for the \gls{bf} formulation.

\subsubsection{Poisson Point Process}
A \gls{ppp} is an \gls{rfs} characterized by an intensity function $\kappa(x)$. The number of elements in the set is Poisson distributed, with mean $\lambda = \int \kappa(x)\,dx$, while the spatial distribution of the points is given by the normalized intensity $c(x) = \kappa(x)/\lambda$, so that $\kappa(x) = \lambda c(x)$ \cite{ristic2013tutorial}.

The density of a \gls{ppp} $X$ is given by
\begin{equation}\label{eq:ppp}
\pi(X) = \exp(-\lambda)
\prod_{x \in X} \kappa(x),
\end{equation}
and the corresponding cardinality distribution is Poisson,
\begin{equation}\label{eq:ppp_cardinality}
    P(|X| = n) = \frac{\lambda^n e^{-\lambda}}{n!}.
\end{equation}
In tracking applications, \glspl{ppp} are commonly used to model clutter (false alarms), sets of undetected targets, and the birth process of new targets.

\subsubsection{Bernoulli RFS}
A Bernoulli \gls{rfs} models the presence of a single element that may or may not exist, \ie, the set $X$ is either empty or contains a single element.
It is characterized by an existence probability $q \in [0,1]$ and a spatial density~$p(x)$ defined on the single element when it exists~\cite{mahler2007statistical}.

The density of a Bernoulli \gls{rfs} $X$ is given by
\begin{equation}
\pi(X) =
\begin{cases}
1 - q, & X = \emptyset, \\
q\,p(x), & X = \{x\}, \\
0, & |X| \geq 2.
\end{cases}
\end{equation}

The Bernoulli \gls{rfs} captures both the uncertainty in target existence and the uncertainty in the target state. This makes it a natural model for single-target tracking problems.

\subsection{Bernoulli Filtering in the RFS Framework}\label{sec:bf_rfs_framework}
In Bernoulli filtering, the target state at time $k$ is modeled as a Bernoulli \gls{rfs} and the measurement set is modeled as an \gls{rfs} $Z_k$. The goal is to recursively estimate the posterior Bernoulli density given the sequence of measurement sets.

The posterior density at time $k-1$ is denoted~$\pi(X_{k-1} \mid Z_{1:k-1})$, where $Z_{1:k-1} = (Z_1, \ldots, Z_{k-1})$ is the sequence of measurement sets up to time $k-1$. 
Similar to the standard Bayesian filter, such as the \gls{kf}, the filtering recursion consists of prediction and update steps \cite{mahler2007statistical}.

The prediction step propagates the density from time~$k-1$ to $k$ using the Chapman--Kolmogorov equation,
\begin{align}\label{eq:prediction}
    &\pi_{k\mid k-1}(X_{k} \mid Z_{1:k-1}) = \nonumber\\
    & \;\;\; \int \! \phi(X_k \mid X_{k-1}) \pi_{k-1\mid k-1}(X_{k-1} \mid Z_{1:k-1}) \,\delta X_{k-1},
\end{align}
where $\phi(X_k \mid X_{k-1})$ is the target transition density, which accounts not only for the motion of an existing target but also for the birth of new targets and the death (disappearance) of existing ones, so that $X_k$ and $X_{k-1}$ may differ in cardinality. 
The integral with respect to $\delta X$ is a set integral in the \gls{fisst}
sense, defined as~\cite{mahler2007statistical}
\begin{align}
    \int f(X)\,\delta X
    =
    \sum_{n=0}^{\infty}\frac{1}{n!}
    \int f(\{x_1,\dots,x_n\})\,dx_1\cdots dx_n .
\end{align}

The update step incorporates the measurement set $Z_k$ using Bayes' rule,
\begin{align}\label{eq:update}
    &\pi_{k\mid k}(X_k \mid Z_{1:k}) = \nonumber\\
    & \qquad \frac{g(Z_k \mid X_k)\,\pi_{k\mid k-1}(X_k \mid Z_{1:k-1})}{\int g(Z_k \mid X_k)\,\pi_{k\mid k-1}(X_k \mid Z_{1:k-1}) \, \delta X_k},
\end{align}
where $g(Z_k \mid X_k)$ is the measurement set likelihood.

For the Bernoulli \gls{rfs}, these equations reduce to recursive updates of the existence probability and the spatial density. The specific form of the update depends on the measurement likelihood model, which is defined in the following section.
\section{Threshold-Dependent Measurement Model}\label{sec:method}
In the Bernoulli \gls{rfs} formulation, the measurement update is determined by the measurement set likelihood~$g(Z \mid X)$. For the single-target case, this reduces to specifying the likelihood of the measurement set $Z$ conditioned on the target state $x$ and the target existence.

In this section, we define a measurement model where the probability of detection depends explicitly on the sensor detection threshold. We consider a general measurement vector consisting of a conventional measurement and a scalar intensity or amplitude measurement. The intensity measurement is subject to thresholding, such that the complete measurement is reported only when the intensity exceeds the detection threshold.

\subsection{Single-Sensor Model}\label{sec:single_sensor_model}
Consider a single sensor observing an existing target with state~$x \in \mathbb{R}^{n_x}$. A reported target-generated measurement is represented by
\begin{equation}
z = (y^T,a)^T \in \mathbb{R}^{n_y+1},
\end{equation}
where $y\in \mathbb{R}^{n_y}$ denotes the conventional, non-intensity part of the measurement, such as a position or range-bearing measurement, and $a$ denotes a scalar intensity or amplitude measurement, such as a \gls{snr} or \gls{rss} measurement. To improve readability, the sensor index~$s$ is omitted here.

The target-generated measurement set~$Z_T$ is either empty or contains a single reported measurement,
\begin{equation}
Z_T =
\begin{cases}
{(y^T,a)^T}, & \text{if } a > \gamma, \\
\emptyset, & \text{otherwise},
\end{cases}
\end{equation}
where $\gamma$ is the detection threshold. Thus, when the intensity measurement does not exceed the threshold, neither $a$ nor the associated measurement $y$ is reported.

The conventional measurement and intensity are modeled according to
\begin{subequations}
    \begin{align}
        y = h_y(x) + e_y, \\
        a = h_a(x) + e_a,
    \end{align}
\end{subequations}
where $h_y(x)$ and $h_a(x)$ are the corresponding measurement functions and $e_y$ and $e_a$ denote measurement noise. In the following, the intensity measurement is assumed to have zero-mean Gaussian noise with variance $\sigma_a^2$, \ie, $e_a \sim \mathcal{N}(0,\sigma_a^2)$. The measurement model for $y$ is left general, allowing for any distribution, including Gaussian or non-Gaussian noise.

The probability of detection is denoted $P_D(x;\gamma)$, emphasizing its dependence on the detection threshold $\gamma$. In general,
\begin{equation}
P_D(x;\gamma) = P_{D,\max} \Pr(a>\gamma\mid x),
\end{equation}
where $P_{D,\max}$ denotes the maximum probability of detection. Under the Gaussian intensity model,
\begin{equation}\label{eq:pd_gaussian}
P_D(x;\gamma) = P_{D,\max}
\left[
    1-\Phi\left(
        \frac{\gamma-h_a(x)}{\sigma_a}
    \right)
\right],
\end{equation}
where $\Phi(\cdot)$ denotes the standard normal \gls{cdf}.

Conditioned on detection, the measurement likelihood can be factorized into the conventional measurement and intensity components. Assuming conditional independence between $y$ and $a$ given $x$, this gives
\begin{subequations}
    \begin{align}
        \ell\left(y,a \mid x,\gamma\right) &= \ell(y\mid x) \, \ell(a\mid x, \gamma) \\
        &= \ell(y\mid x) \, \mathcal{N}_{\text{Tr}}(a;h_a(x),\sigma_a^2,\gamma,\infty),
    \end{align}
\end{subequations}
where the truncated normal distribution is defined as
\begin{equation}\label{eq:trunk_gauss}
\mathcal{N}_{\text{Tr}}(a; h_a(x), \sigma_a^2, \gamma_l, \gamma_u)
=
\frac{
\mathcal{N}(a; h_a(x), \sigma_a^2)
\chi_{(\gamma_l,\gamma_u)}(a)
}{
\int_{\gamma_l}^{\gamma_u}
\mathcal{N}(u; h_a(x), \sigma_a^2)\,du
},
\end{equation}
with $\chi(\cdot)$ being the indicator function, and $\gamma_l$ and $\gamma_u$ denoting the lower and upper truncation limits, respectively.

As illustrated in \Figref{fig:trunk_gauss}, the conditional intensity likelihood is zero below the threshold $\gamma$ and follows a Gaussian distribution above the threshold, normalized by the probability of exceeding the threshold to ensure that it integrates to one over the valid measurement range.
\begin{figure}
    \centering
    \tikzsetnextfilename{trunk_gauss}
    \definecolor{mycolor1}{rgb}{0.00000,0.44706,0.69804}%
\begin{tikzpicture}[scale=1]
        \begin{axis}[
        axis lines=middle,
        xlabel={$a$},
        ylabel={$\ell(a \mid x,\gamma)$},
        xtick={0.5,2,4},
        xticklabels={$\gamma$, 2, 4},
        enlargelimits=0.15,
        samples=200,
        xmin=0,
        xmax=4.5,
        domain=-2:5,
        clip=false,
        every axis y label/.style={at={(axis description cs:0.1,1)},anchor=south},
        every axis x label/.style={at={(axis description cs:1,0.05)},anchor=west},
        line width=1.5pt,
        width=0.8\columnwidth,
        height=0.5\columnwidth,
        ]
        
        % Parameters
        \def\mu{1}
        \def\sigma{0.7}
        \def\gammarel{0.5}
        \def\pD{0.762} % Approximate normalization constant from CDF
        
        % Heaviside function: 1 if x > gamma
        \addplot [
        mycolor1,
        domain=\gammarel:5,
        ]
        {1/(sqrt(2*pi*\sigma^2)) * exp(-((x-\mu)^2)/(2*\sigma^2)) / \pD};

        \addplot [
        mycolor1,
        domain=-0.2:\gammarel,
        ]
        {0};

        \draw[mycolor1] (axis cs:\gammarel,0) -- (axis cs:\gammarel,0.585);
    
        \draw[dashed] (axis cs:\gammarel,0) -- (axis cs:\gammarel,0.8);
        \node at (axis cs:3+1.3,0.8) {$\mathcal{N}_{\text{Tr}}(a; h_a(x), \sigma^2, \gammarel, \infty)$};    
        \end{axis}
\end{tikzpicture}
    \caption{Illustration of the truncated Gaussian likelihood for the intensity measurement~$a$ conditioned on detection, with $h_a(x) = 1$, $\sigma_a = 0.7$, and threshold $\gamma = 0.5$. The dashed line indicates the threshold $\gamma$.}
    \label{fig:trunk_gauss}
\end{figure}

The resulting single-sensor measurement set likelihood, without clutter, is
\begin{align}
g(Z \mid X &= \{x\}, \gamma) = \nonumber\\
&
\begin{cases}
P_D(x;\gamma)\,\ell(y,a \mid x,\gamma),  & Z_T = \{(y^T,a)^T\},\\
1-P_D(x;\gamma), & Z_T = \emptyset.
\end{cases}
\end{align}

\subsection{Clutter Model}\label{sec:method_clutter_model}
To account for false alarms (clutter), the measurement set is modeled as the union of independent target-generated measurements and clutter measurements. Thus, the measurement set is extended as
\begin{equation}
Z = Z_T \cup Z_{\text{FA}},
\end{equation}
where $Z_T \in \{\emptyset,\{z\}\}$ is the target-generated measurement and $Z_{\text{FA}}$ is the set of clutter measurements.

The clutter is modeled as a \gls{ppp} with intensity $\kappa(z)$. The likelihood of a clutter set $Z_{\text{FA}}$ is given by \eqref{eq:ppp}. For a normalized spatial distribution $c(z)$ and expected number of false alarms $\lambda$, the intensity is
\begin{equation}
\kappa(z) = \lambda c(z), \qquad \int c(z)\,dz = 1,
\end{equation}
where $\lambda = \int \kappa(z)\,dz$ is the expected number of clutter measurements.

As with target-generated measurements, clutter measurements are subject to the detection threshold~$\gamma$. However, the threshold only affects the intensity or amplitude component~$a$ of the clutter measurement, while the conventional measurement component~$y$ remains distributed according to its original clutter model. 
Assuming conditional independence of $y$ and $a$,
\begin{equation}
c(z) = c(y,a) = c_y(y)c_a(a),
\end{equation}
where $c_y(y)$ and $c_a(a)$ are the marginal distributions of the conventional measurement and intensity, respectively.
The clutter distribution conditioned on the threshold is therefore
\begin{equation}
c(z\mid\gamma) = c_y(y)c_a(a\mid\gamma),
\end{equation}
where the thresholded intensity distribution is defined as
\begin{equation}
c_a(a\mid\gamma)
=
\frac{
c_a(a)\chi(a>\gamma)
}{
\displaystyle
\int_{\gamma}^{\infty}c_a(u)\,du
}.
\end{equation}
The corresponding thresholded clutter intensity is
\begin{equation}
\kappa(z\mid\gamma) = \lambda c(z\mid\gamma).
\end{equation}
In this formulation, the threshold changes the distribution of the intensity component~$a$ but does not otherwise alter the distribution of the conventional measurement~$y$. 

The expected number of clutter measurements is here assumed to be independent of the detection threshold, but the formulation allows for a threshold-dependent clutter rate $\lambda(\gamma)$ if desired.

\subsection{Measurement Likelihood with Clutter}
Combining the single-sensor likelihood above with the clutter model, the measurement set likelihood given the target state $x$ and threshold $\gamma$ is \cite{ristic2013tutorial}
\begin{equation}\label{eq:combined_likelihood}
\begin{aligned}
g(Z \mid X = \{x\}, \gamma)
&=
\exp(-\lambda)
\Bigg[
\left(1 - P_D\left(x;\gamma\right)\right) \prod_{z \in Z} \kappa(z\mid\gamma) \\
&\hspace{-1cm} + P_D(x;\gamma) \sum_{z \in Z}
\ell(z \mid x,\gamma)
\prod_{\bar{z} \in Z \setminus \{z\}} \kappa(\bar{z} \mid \gamma)
\Bigg].
\end{aligned}
\end{equation}

The first term corresponds to missed detection, where all measurements are clutter, and the second term corresponds to the case where one measurement originates from the target and the remaining measurements are clutter.

For the empty set $Z=\emptyset$, this reduces to
\begin{equation}
g(\emptyset \mid X = \{x\}, \gamma) = \exp(-\lambda)\bigl(1 - P_D(x;\gamma)\bigr).
\end{equation}

The likelihood derived above directly enters the measurement update in the Bernoulli \gls{rfs} framework. In particular, when the target exists, the measurement likelihood is given by $g(Z \mid X = \{x\}, \gamma)$ as defined above. 

In the absence of a target, the measurement set consists purely of clutter, with likelihood
\begin{equation}
g(Z \mid X = \emptyset, \gamma)
=
\exp(-\lambda) \prod_{z \in Z}\kappa(z \mid \gamma).
\end{equation}

This formulation allows for consistent updating of both the target existence probability and the state estimate, while explicitly capturing how the detection probability depends on the threshold $\gamma$.
\section{Recursive Filtering Equations}\label{sec:filter_eqs}
The \gls{bf} consists of a prediction step and an update step that recursively compute the posterior existence probability and spatial density of the target state based on incoming measurements from multiple sensors. Let $\mathcal{Z}_k = \big( Z_k^{(1)}, \ldots, Z_k^{(S)} \big)$ denote all the measurement sets from $S$ sensors at time $k$. The filter equations are derived from the general \gls{rfs} filtering equations presented in \Secref{sec:background}, using the measurement likelihood model defined in \Secref{sec:method}.

\subsection{Prediction Step}
The prediction step from \eqref{eq:prediction} propagates the posterior density from time $k-1$ to $k$ using the target transition density $\phi(X_k \mid X_{k-1})$, which for a Bernoulli \gls{rfs} decomposes into a birth component and a survival component. Death is captured implicitly as the complement of survival: since a target that dies has no state, it contributes no spatial density and only reduces the predicted existence probability. For a Bernoulli \gls{rfs}, the predicted existence probability and spatial density are given by \cite{ristic2013tutorial}:
\begin{subequations}
    \begin{align}
        q_{k\mid k-1} = p_b \cdot  (1 &- q_{k-1\mid k-1}) + p_s q_{k-1\mid k-1}, \\[4pt]
        p_{k\mid k-1}(x_k\mid \mathcal{Z}_{1:k-1}) &= \scriptstyle
        \frac{p_b \cdot (1 - q_{k-1\mid k-1}) b(x_k;\mathcal{Z}_{k-1})}{q_{k\mid k-1}} \nonumber\\
        &\hspace{-2cm} \scriptstyle + \frac{p_s q_{k-1\mid k-1} \int f(x_k \mid x_{k-1}) p_{k-1\mid k-1}(x_{k-1} \mid \mathcal{Z}_{1:k-1}) \,dx_{k-1}}{q_{k\mid k-1}},
    \end{align}
\end{subequations}
where $p_b$ is the birth probability, $p_s$ is the survival probability, $b(x_k;\mathcal{Z}_{k-1})$ is the birth spatial density, and~$f(x_k \mid x_{k-1})$ is the single-target motion density describing how a surviving target's state evolves.

For a linear Gaussian model, as considered in this work, this is given by:
\begin{equation}
f(x_k \mid x_{k-1}) = \mathcal{N}(x_k; F x_{k-1}, Q),
\end{equation}
where $F$ is the state transition matrix and $Q$ is the process noise covariance.

\subsection{Update Step}
Given measurements $\mathcal{Z}_k$ at time $k$, the posterior spatial density is \cite{ristic2013tutorial}:
\begin{equation}
p_{k\mid k}(x_k \mid \mathcal{Z}_{1:k}) = \frac{g(\mathcal{Z}_k \mid \{x_k\}, \gamma) p_{k\mid k-1}(x_k \mid \mathcal{Z}_{1:k-1})}{I_0},
\end{equation}
where $\mathcal{Z}_{1:k} = \big(\mathcal{Z}_1, \ldots, \mathcal{Z}_k\big)$ is the sequence of measurement sets up to time $k$, and
\begin{equation}
I_0 = \int g\left(\mathcal{Z}_k \mid \{x\}, \gamma\right) p_{k\mid k-1}(x \mid \mathcal{Z}_{1:k-1}) \,dx.
\end{equation}
The updated existence probability is:
\begin{equation} 
q_{k\mid k} = \frac{q_{k\mid k-1} I_0}{1 - q_{k\mid k-1} + q_{k\mid k-1} I_0}.
\end{equation}

\subsubsection{Multiple Sensors}
Assuming the sensor measurements are conditionally independent given the target state, the combined measurement likelihood factorizes as
\begin{equation}\label{eq:multisensor_independence}
    g(\mathcal{Z}_k \mid X_k) = \prod_{s=1}^{S} g_s\big(Z_k^{(s)} \mid X_k\big),
\end{equation}
where $g_s(\cdot \mid X_k)$ denotes the single-sensor measurement likelihood from \Secref{sec:method} for sensor $s$. Consequently, a joint update using \eqref{eq:multisensor_independence} is mathematically equivalent to processing the sensors sequentially.

For multi-sensor processing, the update step above is thus applied iteratively for each sensor $s = 1, \ldots, S$. The target existence probability and spatial density are updated sequentially using the measurement set $Z_k^{(s)}$ of each sensor, as defined in \Secref{sec:method}:
\begin{subequations}\label{eq:req_update_full}
    \begin{align} \label{eq:req_ex_prob_update}
        q_{k\mid k} &= \frac{(1-\Delta_k^{(s)}) q_{k\mid k-1}}{1 - \Delta_k^{(s)} q_{k\mid k-1}}, \\
        p_{k\mid k}(x_k) &= \, p_{k\mid k-1}(x_k) \Bigg[ 1 - P_{D,s}(x_k;\gamma) + \nonumber\\ 
        \label{eq:req_density_update}
        & \,
        \sum_{z \in Z^{(s)}_k} \frac{P_{D,s}(x_k;\gamma) \ell^{(s)}(z \mid x_k, \gamma)}{\kappa(z)} \Bigg] \big/ \left[ 1 - \Delta_k^{(s)} \right].
    \end{align}
\end{subequations}
where
\begin{align}\label{eq:delta_k}
\Delta_k^{(s)} &= \overbrace{\int P_{D,s}(x;\gamma) p_{k\mid k}(x) \,dx}^{= I_1} - \nonumber\\
& \quad \sum_{z \in Z^{(s)}_k} \nicefrac{ \overbrace{\int P_{D,s}(x;\gamma) \ell^{(s)}(z \mid x, \gamma) p_{k\mid k}(x) \,dx}^{= I_2(z)}}{\kappa(z)}.
\end{align}

% \subsection{State Estimation}
% Given the posterior distribution, the \gls{mmse} estimate of the target state is:
% \begin{equation}
% \hat{x}_{k\mid k} = \int x \, p_{k\mid k}(x) \,dx \approx \sum_{i=1}^N w^{(i)}_{k\mid k} x^{(i)}_{k\mid k},
% \end{equation}
% which is typically computed only when the existence probability exceeds a threshold: $q_{k\mid k} > q_{\text{threshold}}$.

\subsection{Particle Filter Implementation}
The full \gls{bf} with the measurement model defined in \Secref{sec:method} is implemented using a \gls{pf} approach. The recursive equations for the existence probability and spatial density are approximated using a set of weighted particles. The detailed steps of the \gls{pf} implementation, including resampling and state estimation, is outlined in \Algref{alg:bernoulli_pf} in the Appendix.
Note that the term $\big[ 1 - \Delta_k^{(s)} \big]$ in the spatial density update \eqref{eq:req_density_update} is a normalization factor that ensures the updated density integrates to 1. This is not explicitly computed in the implementation since the density is represented by particles and weights that are normalized after the update step.
% The key parameters of the algorithm are as follows:
% \begin{itemize}
%     \item $N$: Number of particles representing the target state
%     \item $B$: Number of birth particles
%     \item $p_b$: Birth probability of the target
%     \item $p_s$: Survival probability of the target
%     \item $P_{D,s}(x_k;\gamma)$: Probability of detection for sensor $s$ given state $x$ and detection threshold $\gamma$
%     \item $f(x_k \mid x_{k-1})$: State transition probability
%     \item $\ell^{(s)}(z \mid x_k, \gamma)$: Single-target measurement likelihood from sensor $s$
%     \item $b(x_k ; \mathcal{Z}_{k-1})$: Birth density function for new target particles
%     \item $\lambda$: Clutter intensity (expected number of false alarms)
%     \item $c(z)$: Clutter spatial distribution
%     \item $q_{\text{threshold}}$: Existence probability threshold for state estimation
% \end{itemize}
\section{Simulation Setup}\label{sec:sim_model}
A \gls{pf} implementation of the \gls{bf} is evaluated in a simulated 2D tracking scenario with multiple sensors, clutter, and detection uncertainty. The simulation parameters are chosen to reflect realistic tracking conditions while allowing us to analyze the impact of the detection threshold on filter performance.

\subsection{State Space and Dynamics}
The target state is defined in a 2D polar coordinate system, where the state vector includes the bearing and range of the target relative to the sensor, as well as their respective velocities,
$x = [\psi, \dot{\psi}, r, \dot{r}]^T$, where $\psi$ is bearing (rad) and $r$ is range (m).

The dynamics are modeled as two independent \gls{cv} processes for the bearing and range components, each driven by white acceleration noise. The sampling interval is $T = 1$\,s. The state transition matrix $F$ and process noise covariance $Q$ are defined as follows:
\begin{equation}
F =
\left[
\begin{smallmatrix}
1 & T & 0 & 0 \\
0 & 1 & 0 & 0 \\
0 & 0 & 1 & T \\
0 & 0 & 0 & 1
\end{smallmatrix}
\right],
\quad
Q =
\left[
\begin{smallmatrix}
\sigma_\psi^2\frac{T^4}{4} & \sigma_\psi^2\frac{T^3}{2} & 0 & 0 \\
\sigma_\psi^2\frac{T^3}{2} & \sigma_\psi^2 T^2 & 0 & 0 \\
0 & 0 & \sigma_r^2\frac{T^4}{4} & \sigma_r^2\frac{T^3}{2} \\
0 & 0 & \sigma_r^2\frac{T^3}{2} & \sigma_r^2 T^2
\end{smallmatrix}
\right]
\end{equation}
where $\sigma_\psi$ and $\sigma_r$ denote the standard deviations of the angular and radial acceleration process noise, respectively. For the simulations, $\sigma_\psi=0.05$\,rad/s$^2$ and $\sigma_r=1$\,m/s$^2$, reflecting moderate target maneuvering.

\subsection{Sensor Configuration}\label{sec:sensor_config}
In the simulation, we consider a scalar \gls{rss} measurement of the target, \ie, only the intensity/amplitude component $a$ of the general measurement model introduced in \Secref{sec:method}, with no separate conventional measurement ($n_y=0$). We consider a scenario involving multiple sensors with nonlinear measurement functions and directional-sensitivity patterns.

Specifically, we simulate $S = 4$ sensors placed at the origin, with their orientations separated by $90^\circ$. The measurement function of each sensor depends on both range and bearing, where the bearing component is nonlinear due to the directional sensitivity of the sensor.

The measurement model for sensor $s$ with orientation $\phi^{(s)}$ is:
\begin{align}\label{eq:sim_meas_model}
h^{(s)}(x) &=
-50 -20 \cdot \log_{10}(r)
+ h_\psi\left(\psi + \phi^{(s)}\right)
\end{align}
where $\phi^{(s)} = (s-1) \cdot \frac{\pi}{2}$ for $s=1,2,3,4$ corresponds to the sensor orientations at 0$^{\circ}$, 90$^{\circ}$, 180$^{\circ}$, and 270$^{\circ}$, respectively. The range component is modeled using a logarithmic relationship based on the \gls{rss}. In contrast, the bearing component is described by the nonlinear function $h_\psi(\cdot)$, which characterizes the directional sensitivity of the sensor, designed to resemble a realistic directional antenna pattern from \cite{zetterqvist2025misseddetection}.

The directional-sensitivity pattern is modeled using a \gls{fs} with $M$ harmonics, given by
\begin{equation}\label{eq:fs_model}
h_\psi(\psi + \phi^{(s)}) = \sum_{m=-M}^{M} c_m\, e^{i m (\psi + \phi^{(s)})},
\end{equation}
where $c_m$ denotes the complex Fourier coefficient corresponding to harmonic $m$. The \gls{fs} provides a natural and compact representation of the directional-sensitivity pattern, capturing its periodic dependence on the bearing $\psi$ while allowing the response to be represented using a finite number of harmonics.
The resulting measurement model for the bearing component $h_\psi(\cdot)$ is shown in \Figref{fig:measurement_model}.
\begin{figure}[tb]
    \centering
    \tikzsetnextfilename{measurement_model_scalar_dbm}
    \input{Tikz/measurement_model_scalar_dbm.tex}
    \caption{Measurement model for the sensors showing the nonlinear relationship between the target state and the measurements for the bearing component.}
    \label{fig:measurement_model}
\end{figure}

Under this model, the detection probability for sensor~$s$ is defined as a function of the target state and the detection threshold $\gamma$. It consists of two components: a constant maximum detection probability $P_{D,\max}^{(s)}$ that represents the best-case detection performance, and a state- and threshold-dependent component that models the degradation in detection performance as the target's signal strength approaches the detection threshold. Following the same form as \eqref{eq:pd_gaussian}, the detection probability for sensor $s$ is:
\begin{equation}
P_{D,s}(x; \gamma^{(s)}) = P_{D,\max}^{(s)}
\left[
    1-\Phi\left(
        \frac{\gamma^{(s)}-h^{(s)}(x)}{\sigma^{(s)}}
    \right)
\right],
\end{equation}
where $\Phi(\cdot)$ is the \gls{cdf} of the standard normal distribution, and $\gamma^{(s)}$ is the detection threshold parameter for sensor~$s$. The maximum detection probability $P_{D,\max}^{(s)}$ is set to 0.9 for all sensors, reflecting the maximum achievable detection performance under ideal conditions.
The measurement noise is assumed to be Gaussian with per-sensor variance ${\sigma^{(s)}}^2 = \sigma^2$ for all sensors, where $\sigma$ is set to 1\,dBm to reflect realistic measurement noise levels in a radio signal tracking scenario~\cite{Ramirez2021BLE}.

Although the formulation allows for sensor-specific thresholds $\gamma^{(s)}$, a common threshold $\gamma = -95$\,dBm is used for all sensors in the simulations to simplify the analysis and focus on the impact of the detection threshold on tracking performance. The chosen threshold value is representative of typical detection thresholds in radio signal tracking applications, where signals below this level are likely to be indistinguishable from noise, leading to missed detections \cite{nrf52840v1_11, Ramirez2021BLE}.

The measurement likelihood for sensor $s$ is:
\begin{equation}
\ell^{(s)}(z \mid x, \gamma) = \mathcal{N}_{\text{Tr}}(z; h^{(s)}(x), \sigma^2, \gamma, \infty),
\end{equation}
consistent with the single-sensor intensity likelihood of \Secref{sec:method}.

\subsection{Clutter Model}
The clutter is modeled as a \gls{ppp} with intensity $\lambda$ and spatial distribution $c(z)$. The clutter intensity~$\lambda$ represents the expected number of false alarms per scan. For our simulations, $\lambda = 2$ is chosen to reflect a moderate clutter environment. The spatial distribution~$c(z)$ is defined as a truncated Gaussian centered below the detection threshold to reflect the fact that false alarms are more likely to occur near the detection threshold due to noise fluctuations.

The clutter spatial distribution is defined as:
\begin{equation}\label{eq:sim_clutter}
c(z\mid\gamma) = \mathcal{N}_{\text{Tr}}(z; \mu_c, \sigma_c^2, \gamma, \infty) ,
\end{equation}
where $\mu_c$ is set to -98\,dBm and $\sigma_c$ is set to 2\,dBm, so that reported clutter measurements are concentrated just above the detection threshold $\gamma$, consistent with false alarms arising from noise fluctuations that occasionally cross the threshold.

\subsection{Target Trajectory}
The target follows a trajectory defined by the state transition model, the initial state, and the process noise. The target's initial state is set to $x_0 = [\psi_0, \dot{\psi}_0, r_0, \dot{r}_0]^T$, where~$\psi_0$ and~$r_0$ are the initial bearing and range, and~$\dot{\psi}_0$ and~$\dot{r}_0$ are the initial velocities in bearing and range, respectively. The target's trajectory is generated by propagating the state through the dynamic model while adding process noise at each time step. The resulting trajectory is illustrated in \Figref{fig:trajectory_and_sensors}, showing the target's path relative to the sensor detection regions. In total, the simulation runs for $K = 80$ time steps. The target appears at time step 10 and disappears at time step 70, allowing us to evaluate the filter's ability to detect and track the target over time while accounting for missed detections and false alarms. 

The trajectory is designed to pass through the detection regions of multiple sensors, providing a realistic scenario for evaluating the \gls{bf}'s performance in a multi-sensor tracking context. The sensors' surveillance areas are depicted as shaded regions in \Figref{fig:trajectory_and_sensors}, illustrating the areas where the target can be detected based on the measurement model and detection threshold, set to $\gamma = -95$\,dBm for the simulations.
\begin{figure}[tb]
    \centering    
    \tikzsetnextfilename{surveillance_region_dbm}
    \input{Tikz/surveillance_region_dbm.tex}
    \caption{True target trajectory (black) and sensor surveillance areas (shaded regions) for the simulated scenario. The target appears at time step 10 and disappears at time step 70.
    The sensors are located at the origin with orientations at 0$^\circ$, 90$^\circ$, 180$^\circ$, and 270$^\circ$, respectively. The shaded regions indicate the areas where the target can be detected based on the measurement model and detection threshold $\gamma = -95$\,dBm.
    }
    \label{fig:trajectory_and_sensors}
\end{figure}

\subsection{Filter Parameters}
The \gls{pf} implementation of the \gls{bf} is configured with the parameters outlined in \Tabref{tab:filter_parameters}. The number of particles for survival and birth is selected to balance computational complexity and estimation accuracy. The birth and survival probabilities are determined using a grid search over candidate values of $p_b$ and $p_s$. Specifically, $p_s$ is varied from $0.8$ to $0.99$ in increments of $0.01$, while $p_b$ is varied from $0.001$ to $0.1$ in increments of $0.005$. The combination yielding the lowest \gls{gospa} metric is selected. The initial existence probability is set to zero, reflecting the assumption that the target does not exist at the start of the simulation, while the existence threshold is set to $0.5$, meaning that the filter considers the target to exist when the existence probability exceeds this threshold.

% the following parameters, where the birth and survival probabilities are selected based on a grid search to optimize performance in terms of the \gls{gospa} metric. The filter parameters are summarized in \Tabref{tab:filter_parameters}.
% The grid search results for birth probability $p_b$ and survival probability $p_s$ are shown in \Figref{fig:parameter_grid_search}, where the average \gls{gospa} metric across 5 \gls{mc} trials is plotted for different parameter combinations. The selected parameters are highlighted with a white star, indicating the optimal choice based on the grid search.
% \begin{figure}[tb]
%     \centering
%     \subfloat[$P_D$ with $\gamma = -95$\,dBm]{%
%         \includegraphics[width=0.95\columnwidth]{sweep_ps_pb_results.png}
%         \label{fig:ps_pb_sweep}
%     }
    
%     \vspace{0.5em}
    
%     \subfloat[Constant $P_D$]{%
%         \includegraphics[width=0.95\columnwidth]{sweep_ps_pb_results_const.png}
%         \label{fig:ps_pb_const}
%     }
    
%     \caption{Grid search results for birth probability $p_b$ and survival probability $p_s$ showing the average \gls{gospa} metric across 5 \gls{mc} trials. The selected parameters are highlighted with a white star.}
%     \label{fig:parameter_grid_search}
% \end{figure}
\begin{table}[tb]
    \centering
    \caption{Parameters of the \gls{pf} implementation of the \gls{bf}.}
    \begin{tabular}{lc}
        \toprule
        Parameter & Value \\
        \midrule
        Survival particles ($N$) & 1000 \\
        Birth particles ($B$) & 500 \\
        Birth probability ($p_b$) & 0.01 \\
        Survival probability ($p_s$) & 0.9 \\
        Initial existence probability ($q_{0|0}$) & 0 \\
        Existence threshold ($q_{\text{threshold}}$) & 0.5 \\
        \bottomrule
    \end{tabular}
    \label{tab:filter_parameters}
\end{table}

\subsection{Birth Particle Initialization}
Birth particles are initialized using either uniform or measurement-driven initialization, depending on the availability of measurements.

If no measurements are available from a randomly selected sensor, the position components of the birth particles, $\psi$ and $r$, are initialized uniformly over the state space, allowing the filter to explore all possible target states without bias. This approach is suitable for scenarios where there is no prior information about the target's location or when the target is expected to appear randomly within the surveillance area~\cite{ristic2012bernoulli},
\begin{equation}
b_k(\psi, r) = \text{Uniform}(\psi \in [-\pi, \pi], r \in [0, 100]).
\end{equation}

When measurements are available, a measurement-driven initialization strategy is employed. 
The birth particles are initialized around a randomly selected measurement from a randomly chosen sensor.
Since the measurement function is nonlinear in bearing, a random angle $\psi$ is drawn uniformly from $[-\pi, \pi]$ and the corresponding range $r$ is computed based on the measurement and the sensor's orientation using \eqref{eq:sim_meas_model}. 
This approach allows us to generate birth particles that are more likely to be near the true target state \cite{ristic2012bernoulli}.

In both cases, the velocity components $\dot{\psi}$ and $\dot{r}$ are initialized with zero mean and variance according to expected target dynamics from the process noise covariance $Q$. The full birth particle initialization procedure is outlined in \Algref{alg:birth_init}.

\begin{algorithm}[tb]
\caption{Birth Particle Initialization}
\label{alg:birth_init}
\begin{algorithmic}[1]
\For{$i=N+1$ to $N+B$}
    \State Select random sensor $s'$
    % If empty, fall back to uniform initialization
    \If{$Z^{(s')}_k = \emptyset$}
        \State $\!\!(\psi^{(i)}, r^{(i)}) \! \sim \! \text{Uniform}\big(\psi \! \in \! [-\pi,\pi],\, r \! \in \! [0,100]\big)$
        \State \textbf{continue}
    \Else
        \State Select random measurement $z^{(s')}$ from $Z^{(s')}_k$
        \State Draw $\psi^{(i)} \sim \text{Uniform}(-\pi,\pi)$
        \State Compute range $r^{(i)}$ given $\psi^{(i)}$ and  $z^{(s')}$:
        \begin{equation}
        r^{(i)} = 10^{\frac{-\left( z^{(s')} - h_\psi\left(\psi^{(i)} + \phi^{(s')}\right) + 50 \right)}{20}}
        \end{equation}
    % \State Set $x^{(i)}_k = [\psi^{(i)}, 0, r^{(i)}, 0]^T$
    \EndIf
    \State Draw process noise $w \sim \mathcal{N}(\mathbf{0}, Q)$
    \State $\dot{\psi}^{(i)} \gets w_2$, \quad $\dot{r}^{(i)} \gets w_4$
    \State Set $x^{(i)}_k = [\psi^{(i)}, \dot{\psi}^{(i)}, r^{(i)}, \dot{r}^{(i)}]^T$
\EndFor
\end{algorithmic}
\end{algorithm}

\subsection{Performance Metrics}
The performance of the \gls{bf} is evaluated using the \gls{gospa} metric, which provides a comprehensive measure of tracking performance by accounting for localization errors, missed detections, and false alarms \cite{Rahmathullah2017GOSPA}. The \gls{gospa} metric is particularly suitable for evaluating the performance of the \gls{bf} in scenarios with detection uncertainty and clutter, as it captures the trade-offs between different types of errors in a single metric.

The \gls{gospa} metric is developed for multi-target tracking scenarios, and with $c>0$, $0 < \alpha \leq 2$, $1 \leq p < \infty$, and $|X| \leq |\hat{X}|$ it is defined as \cite{Rahmathullah2017GOSPA}:
\begin{align}
    &d_p^{(c,\alpha)}(X, \hat{X}) = \nonumber\\
    &\; \min_{\pi \in \Pi_{|\hat{X}|}} \left( \sum_{i=1}^{|X|} d^{(c)}(x_i, \hat{x}_{\pi(i)})^p + \frac{c^p}{\alpha} (|\hat{X}| - |X|) \right)^{1/p},
\end{align}
where $X$ is the set of true target states, $\hat{X}$ is the set of estimated target states, $d^{(c)}(x, \hat{x}) = \min(c, d(x, \hat{x}))$ is the cut-off distance between a true state and an estimated state, and $\Pi_{|\hat{X}|}$ is the set of all permutations of the indices of the estimated states. If $|X| > |\hat{X}|$, the \gls{gospa} is defined as $d_p^{(c,\alpha)}(X, \hat{X}) = d_p^{(c,\alpha)}(\hat{X}, X)$. 

In the single-target case, where $|X| \leq 1$ and $|\hat{X}| \leq 1$, the \gls{gospa} metric simplifies significantly, as there are at most one true state and one estimated state to compare. In this case, the assignment problem reduces to a simple comparison between the two sets, which can be empty or contain a single element. This allows for a closed-form expression of the \gls{gospa} metric that directly captures the localization error, missed detection error, and false alarm error without needing to solve a combinatorial assignment problem.

The \gls{gospa} for the single-target case can be expressed as:
\begin{equation}\label{eq:gospa_single_target}
d_p^{(c,\alpha)}(X,\hat{X}) =
\begin{cases}
0, & |X| = |\hat{X}| = 0, \\[4pt]
\dfrac{c}{\alpha^{1/p}}, & |X| \neq |\hat{X}|, \\[6pt]
d^{(c)}(\mathbf{x},\hat{\mathbf{x}}), & |X| = |\hat{X}| = 1.
\end{cases}
\end{equation}
The specific error components captured by the \gls{gospa} in the single-target case are:
\begin{itemize}
    \item \textbf{Localization error:} Error in estimated target position when correctly detected
    \item \textbf{Missed detection error:} Penalty for failing to detect an existing target
    \item \textbf{False alarm error:} Penalty for incorrectly declaring target existence
\end{itemize}

For our simulations, we set the \gls{gospa} parameters to $p=2$,~$c=50$\,m, and $\alpha=2$.
These values enable us to effectively capture the trade-offs between localization errors, missed detections, and false alarms when evaluating the performance of the \gls{bf} under different detection threshold settings.

The simulation consists of 50 \gls{mc} trials for each detection threshold value $\gamma$ to assess statistical performance. The results are evaluated in terms of the average \gls{gospa} metric across trials, as well as the existence probability and measurement statistics over time to provide insight into the filter's behavior under different detection conditions.

\subsection{Compared Filters}\label{sec:compared_filters}
Three filter configurations are compared in \Secref{sec:sim_results}, denoted \textbf{Proposed}, \textbf{Const $P_D$}, and \textbf{Unknown $P_D$}.

The \textbf{Proposed} filter uses the threshold-dependent detection probability $P_{D,s}(x;\gamma)$ defined above, with the assumed threshold $\gamma$ swept over a range of values to evaluate the filter's sensitivity to threshold misspecification, $\gamma = \gamma^* = -95$\,dBm recovers the true detection threshold and gives the best achievable performance.

The \textbf{Const $P_D$} baseline instead assumes a constant detection probability $P_D = 0.9$ for all states, independent of $\gamma$. However, the measurement likelihood remains threshold-truncated $\ell^{(s)}(z\mid x,\gamma)$ from \Secref{sec:method}, evaluated at the threshold $\gamma$ assumed by the filter, which is swept over the same range of values as the Proposed filter. This baseline represents a common approach in the literature where the detection probability is treated as a fixed parameter, ignoring its dependence on the target state and detection threshold.

The \textbf{Unknown $P_D$} baseline requires no model of how the detection threshold derives into a state-dependent detection probability $P_{D,s}(x;\gamma)$, and instead estimates each sensor's detection probability online from the measurements, using a recursively updated Beta-distributed belief following the estimator of \cite{li2021robustpmbm}, adapted here to the single-target \gls{bf} and detailed in \Appref{sec:unknown_pd_appendix}.
The measurement likelihood remains the threshold-truncated $\ell^{(s)}(z\mid x,\gamma)$ from \Secref{sec:method}, evaluated over the same range of assumed thresholds $\gamma$ as swept by the Proposed filter and Const $P_D$ baseline.
\section{Simulation Results}\label{sec:sim_results}
\begin{figure}[tb]
    \centering
    \tikzsetnextfilename{baseline_vs_gamma_comparison_scalar_dbm}
    \input{Tikz/baseline_vs_gamma_comparison_scalar_dbm.tex}
    \caption{Performance comparison across different detection thresholds $\gamma$ in $P_{D,s}(x; \gamma)$.
    \textbf{Top:} Total \gls{gospa} metric vs. threshold.
    \textbf{Bottom:} \Gls{gospa} decomposition into localization (blue), missed detection (orange), and false alarm (yellow) components.
    The dashed lines indicate the Const $P_D$ baseline, the dash-dotted lines indicate the Unknown $P_D$ baseline, the shaded regions represent the 95\% \gls{ci} across \gls{mc} trials, and the vertical dashed line illustrates the true threshold value $\gamma^* = -95$\,dBm.}
    \label{fig:results}
\end{figure}

Measurements are simulated using a fixed true detection threshold $\gamma^* = -95$\,dBm. In the \gls{pf} implementations of the \gls{bf}, however, the assumed detection threshold~$\gamma$ in $P_{D,s}(x;\gamma)$ and the likelihood $\ell^{(s)}(z\mid x,\gamma)$ is varied in order to analyze the sensitivity of the filter performance to threshold mismatch. The results are presented in terms of the \gls{gospa} metric and its decomposition, as well as detection statistics and existence probability.

\Figref{fig:results} shows the performance of the \gls{pf} implementation of the \gls{bf} for different assumed detection thresholds~$\gamma$ in~$P_{D,s}(x; \gamma)$.
The threshold determines the detection probability and therefore the amount of information available to the filter. Different threshold values lead to different trade-offs between localization error, missed detections, and false alarms in the \gls{gospa} metric.

For very low threshold values, the detection probability becomes high across large parts of the state space. In this case the filter expects detections from multiple sensors. However, due to the directional-sensitivity patterns and the trajectory in \Figref{fig:trajectory_and_sensors}, the target is usually observed by only a subset of the sensors. When the remaining sensors do not report measurements, the Bernoulli update interprets these missed detections as evidence against the target hypothesis, reducing the estimated existence probability. As a result, the filter may temporarily fail to declare the target, and missed detections become a significant contributor to the \gls{gospa} metric.
\begin{figure}[tb]
    \centering
    \tikzsetnextfilename{gospa_and_existence_time_series_scalar_dbm}
    \input{Tikz/time_series_comparison_scalar_dbm.tex}
    \caption{Mean \gls{gospa} metric (top) and existence probability (bottom) over time for all \gls{mc} trials with detection threshold $\gamma = -95$\,dBm. The proposed approach is shown in solid lines, with the Const $P_D$ baseline in dashed lines and the Unknown $P_D$ baseline in dash-dotted lines. The shaded regions indicate the 95\% \gls{ci} across trials. The black line indicates the target presence.}
    \label{fig:GOSPA_Existence_over_time}
\end{figure}

For threshold values near the true threshold ($\gamma^* = -95$\,dBm), the filter achieves the best performance. In this region the detection probability matches the actual sensing conditions, allowing the filter to correctly interpret both detections and non-detections. The existence probability remains stable when the target is present, and the resulting \gls{gospa} values are significantly lower than those obtained with both the Const $P_D$ and Unknown $P_D$ baselines, as shown in \Figref{fig:results}. The proposed approach achieves a $53.9\%$ lower \gls{gospa} metric at $\gamma^*$ than the Const $P_D$ baseline, and a $37.7\%$ lower \gls{gospa} metric than the Unknown $P_D$ baseline with $k_\beta = 1.1$, and a $35.7\%$ lower \gls{gospa} metric than the Unknown $P_D$ baseline with $k_\beta = 2.0$.

For higher threshold values the detection probability decreases and the target is detected less frequently. This increases the missed detection component of the \gls{gospa} metric, as the filter does not believe target-generated measurements are likely to occur, and thinks that most measurements are clutter. The existence probability decreases accordingly, and the filter may fail to declare the target for extended periods of time, leading to poor tracking performance.

In \Figref{fig:GOSPA_Existence_over_time} the \gls{gospa} metric and existence probability are shown over time for all \gls{mc} trials for the proposed filter compared to all the baselines at $\gamma=\gamma^*$.
The time-series results provide additional insight into the filter's behavior. When the target appears, the existence probability increases rapidly as measurements are received. During the time interval where the target is present, the existence probability usually remains above the estimation threshold, allowing the filter to produce state estimates. When the target disappears, the existence probability decreases accordingly.

\begin{figure}[tb]
    \centering
    \tikzsetnextfilename{measurement_statistics_scalar_dbm}
    \input{Tikz/measurement_statistics_scalar_dbm.tex}
    \caption{Mean number of measurements from all sensors across time for all \gls{mc} trials. All measurements are shown in solid black line, while the target-generated measurements are shown in color (one color per sensor). The black dotted line indicates all true target-generated measurements. The shaded regions indicate the 95\% \gls{ci} across trials. 
    % Since the clutter process intensity is $\lambda = 2$ per sensor and $S=4$ sensors are used, the expected number of clutter measurements is approximately $8$ per time step, which is consistent with the observed statistics.
    }
    \label{fig:measurements_over_time}
\end{figure}
The drops in existence probability around time step~51 and time step~58 correspond to the target being on the edge of the sensor coverage area, as seen in \Figref{fig:trajectory_and_sensors}. In this region the detection probability is lower, and the filter receives fewer detections, which reduces the existence probability. However, the existence probability does not drop to zero due to the prior information and the presence of clutter measurements that provide some evidence of target existence. When the target reappears within sensor coverage, the existence probability increases again, allowing the filter to produce estimates. The \gls{gospa} metric follows a similar pattern, with increased values during periods of low existence probability and improved performance when the existence probability is high.

The proposed filter consistently outperforms both the Const $P_D$ and Unknown $P_D$ baselines across all time steps, keeping the existence probability above the detection threshold when the target is present and achieving lower \gls{gospa} values. The Const $P_D$ baseline struggles to maintain a high existence probability during periods of low detection probability, while the Unknown $P_D$ baseline adapts online but still underperforms compared to the proposed filter that explicitly models the detection threshold.

The measurements from all sensors across time are shown in \Figref{fig:measurements_over_time}, illustrating the clutter and target-generated measurements.
The measurement statistics show that most detections originate from a single sensor at a given time step, which is consistent with the directional-sensitivity patterns and the simulated trajectory. 
Since the clutter intensity is $\lambda = 2$ per sensor and~$S=4$ sensors are used, the expected number of clutter measurements is $8$ per time step, which is consistent with the observed statistics.

\begin{figure*}[tb]
    \centering
    \tikzsetnextfilename{BernoulliPF_scalar_trajectory_dbm}
    \input{Tikz/BernoulliPF_scalar_trajectory_dbm.tex}
    \caption{Single \gls{mc} run with the true detection threshold $\gamma = -95$\,dBm. \textbf{Top left:} True target trajectory (black) and estimated trajectory (dotted light blue). 
    \textbf{Top mid:}  Filter existence probability over time, indicating confidence in target presence. 
    \textbf{Top right:} \Gls{gospa} metric over time, showing tracking performance.
    \textbf{Bottom left:}    
    Measurements from all sensors, showing clutter (dark-colored crosses) and target-generated measurements (colored circles), one color per sensor. 
    \textbf{Bottom mid:} Number of measurements per time step, indicating clutter density, and target detections.
    \textbf{Bottom right:} \Gls{gospa} decomposition over time, showing localization error (blue), missed detection error (orange), and false alarm error (yellow).}
    \label{fig:single_run}
\end{figure*}

For a single \gls{mc} run with $\gamma = -95$\,dBm, \Figref{fig:single_run} illustrates the estimated target trajectory, measurements, and filter estimates over time.
In the single-run example, a few missed detections occur when the target is observed by only one sensor and that sensor also misses the detection. However, the existence probability remains relatively high due to the prior information and the presence of clutter measurements that provide some evidence of target existence. The filter quickly recovers when the target is detected again. This is especially evident around time step~60, where the target is detected again after a period of low existence probability, leading to a rapid increase in existence probability and improved \gls{gospa} performance.

This is also the case when the target is located outside the coverage area of all sensors (time steps 65--68), where the existence probability decreases but does not drop to zero due to the prior and clutter measurements. When the target reappears within sensor coverage, the existence probability increases again at time step~70, allowing the filter to produce a final estimate of the target's state.

Overall, the results demonstrate that explicitly modeling the detection threshold and its impact on the probability of detection improves the performance of the \gls{bf} compared to both the Const $P_D$ and Unknown $P_D$ baselines. As shown in \Figref{fig:results}, the proposed approach achieves a $53.9\%$ lower \gls{gospa} metric at the optimal threshold value than the Const $P_D$ baseline, and, more importantly, a $35.7\%$ lower \gls{gospa} metric than the Unknown $P_D$ baseline, which itself already estimates detection probability online and requires no model of detection probability $P_{D,s}(x;\gamma)$. By better matching the detection model to the actual sensing conditions, the filter can more accurately interpret detections and missed detections, leading to more reliable existence probability estimates and improved state estimation performance. This results in a better balance between localization error, missed detections, and false alarms, especially in regions with limited sensor coverage and varying detection conditions. The Unknown $P_D$ baseline itself still improves substantially over the Const $P_D$ baseline, confirming that online estimation of the detection probability is a useful fallback, especially when the detection threshold is unknown, but explicitly modeling the threshold, when it is known, remains clearly beneficial.

% \section{Discussion}\label{sec:discussion}

\section{Conclusion}\label{sec:conclusion}\glsresetall
This paper introduced a \gls{pf} implementation of the \gls{bf} that explicitly incorporates the detection threshold into the measurement likelihood. The formulation accounts for thresholded measurements, clutter modeled as a \gls{ppp}, and uncertainty in target existence.

The \gls{bf} was implemented using a \gls{pf} and evaluated in a simulated multi-sensor tracking scenario with nonlinear \gls{rss} measurements. The results demonstrate that incorporating the detection threshold into the measurement model can improve tracking performance compared with a baseline approach using a constant detection probability as well as the unknown-detection-probability estimator of \cite{li2021robustpmbm}. The proposed method also provides a more accurate estimate of the target's existence probability, which is crucial for decision-making in tracking applications.

Future work includes extending the framework to multi-target scenarios and investigating adaptive threshold estimation techniques to further enhance performance in dynamic environments.

\section{Acknowledgment}
 
The authors acknowledge the use of generative \gls{ai} tools during the writing of this manuscript.
Specifically, OpenAI's ChatGPT was used to improve grammar, clarity, and readability in \Secref{sec:intro}, and the conclusions of \Secref{sec:sim_results}. GitHub Copilot was used to assist with \LaTeX-formatting of mathematical expressions in the Theoretical Background and Methodology sections, \Secref{sec:background}-\ref{sec:sim_model}, as well as to help with the formatting in \Algref{alg:bernoulli_pf}.
Anthropic's Claude was used to assist with the implementation of the unknown detection probability baseline in \Secref{sec:sim_model}.

The authors carefully reviewed and edited all content generated by \gls{ai} tools to ensure accuracy and coherence with the scientific contributions of the work.
The authors take full responsibility for the content of this paper and confirm that the use of \gls{ai} tools did not influence the scientific integrity or conclusions presented in this manuscript.

% \newpage
\bibliography{refs}

@article{vo2012multi,
  author={Vo, Ba Tuong and See, Chong Meng and Ma, Ning and Ng, Wee Teck},
  journal={IEEE Transactions on Aerospace and Electronic Systems}, 
  title={Multi-Sensor Joint Detection and Tracking with the {Bernoulli} Filter}, 
  year={2012},
  volume={48},
  number={2},
  pages={1385-1402},
  doi={10.1109/TAES.2012.6178069}}

@article{kim2021bernoulli,
  author={Kim, Du Yong and Ristic, Branko and Guan, Robin and Rosenberg, Luke},
  journal={IEEE Transactions on Aerospace and Electronic Systems}, 
  title={A {Bernoulli} Track-Before-Detect Filter for Interacting Targets in Maritime Radar}, 
  year={2021},
  volume={57},
  number={3},
  pages={1981-1991},
  doi={10.1109/TAES.2021.3054715}}

@article{ristic2012bernoulli,
  author={Ristic, Branko and Arulampalam, Sanjeev},
  journal={IEEE Transactions on Aerospace and Electronic Systems}, 
  title={Bernoulli Particle Filter with Observer Control for Bearings-Only Tracking in Clutter}, 
  year={2012},
  volume={48},
  number={3},
  pages={2405-2415},
  doi={10.1109/TAES.2012.6237599}}

@article{ristic2013tutorial,
  author={Ristic, Branko and Vo, Ba-Tuong and Vo, Ba-Ngu and Farina, Alfonso},
  journal={IEEE Transactions on Signal Processing}, 
  title={A Tutorial on {Bernoulli} Filters: Theory, Implementation and Applications}, 
  year={2013},
  volume={61},
  number={13},
  pages={3406-3430},
  doi={10.1109/TSP.2013.2257765}}

@article{zetterqvist2025misseddetection,
  author={Zetterqvist, Gustav and Gustafsson, Fredrik and Hendeby, Gustaf},
  journal={IEEE Sensors Journal}, 
  title={Utilizing Missed Detections in Directional Sensitivity-Based {DOA} Estimation}, 
  year={2026},
  volume={26},
  number={16},
  pages={24299-24310},
  doi={10.1109/JSEN.2026.3705098}}

@ARTICLE{zetterqvist2025DirectionalSensitivity,
  author={Zetterqvist, Gustav and Gustafsson, Fredrik and Hendeby, Gustaf},
  journal={IEEE Sensors Journal}, 
  title={Directional Sensitivity-Based {DOA} Estimation Using a {Fourier} Series Model}, 
  year={2025},
  volume={25},
  number={20},
  pages={38359-38370},
  doi={10.1109/JSEN.2025.3604893}}

@INPROCEEDINGS{Rahmathullah2017GOSPA,
 author={Rahmathullah, Abu Sajana and Garc{\'i}a-Fern{\'a}ndez, {\'A}ngel F. and Svensson, Lennart},
  booktitle={2017 20th International Conference on Information Fusion (Fusion)}, 
  title={Generalized optimal sub-pattern assignment metric}, 
  year={2017},
  volume={},
  number={},
  pages={1-8},
  doi={10.23919/ICIF.2017.8009645}}

@book{mahler2007statistical,
  title={Statistical Multisource-Multitarget Information Fusion},
  author={Mahler, Ronald P. S.},
  year={2007},
  address={Norwood, MA},
  publisher={Artech House}
}

@article{papi2014bernoulli,
  author    = {Francesco Papi and Vladimir Kyovtorov and Raimondo Giuliani and Franco Oliveri and Dario Tarchi},
  title     = {Bernoulli Filter for Track-Before-Detect Using {MIMO} Radar},
  journal   = {IEEE Signal Processing Letters},
  volume    = {21},
  number    = {9},
  pages     = {1145--1149},
  year      = {2014},
  doi       = {10.1109/LSP.2014.2325566}
}

@article{saucan2017multisensor,
  author    = {Augustin-Alexandru Saucan and Mark Coates and Michael Rabbat},
  title     = {A Multisensor Multi-{Bernoulli} Filter}, 
  journal   = {IEEE Transactions on Signal Processing},
  volume    = {65},
  number    = {20},
  pages     = {5495--5509},
  year      = {2017},
  doi       = {10.1109/TSP.2017.2723348}
}

@article{zhang2018joint,
  author    = {Guangpu Zhang and Ce Zheng and Sibo Sun and Guolong Liang and Yifeng Zhang},
  title     = {Joint Detection and {DOA} Tracking with a {Bernoulli} Filter Based on Information Theoretic Criteria},
  journal   = {Sensors},
  volume    = {18},
  number    = {10},
  pages     = {3473},
  year      = {2018},
  doi       = {10.3390/s18103473}
}

@article{garcia2016lmbtbd,
  author    = {{\'A}ngel F. Garc{\'i}a-Fern{\'a}ndez},
  title     = {Track-Before-Detect Labelled Multi-{Bernoulli} Particle Filter with Label Switching},
  journal   = {IEEE Transactions on Aerospace and Electronic Systems},
  volume    = {52},
  number    = {5},
  pages     = {2123--2138},
  year      = {2016},
  doi       = {10.1109/TAES.2016.150343}
}

@article{li2021robustpmbm,
  author={Li, Guchong and Kong, Lingjiang and Yi, Wei and Li, Xiaolong},
  title={Robust Poisson Multi-{Bernoulli} Mixture Filter With Unknown Detection Probability}, 
  journal={IEEE Transactions on Vehicular Technology}, 
  year={2021},
  volume={70},
  number={1},
  pages={886-899},
  doi={10.1109/TVT.2020.3047107}}

@article{houssineau2020possibilistic,
  author = {Branko Ristic and Jeremie Houssineau and Sanjeev Arulampalam},
  title     = {Target Tracking in the Framework of Possibility Theory: The Possibilistic {Bernoulli} Filter},
  journal   = {Information Fusion},
  volume    = {62},
  pages     = {81--88},
  year      = {2020},
  doi       = {10.1016/j.inffus.2020.04.008}
}

@manual{nrf52840v1_11,
  title        = {nRF52840 Product Specification v1.11},
  author       = {{Nordic Semiconductor ASA}},
  year         = {2024},
  organization = {Nordic Semiconductor},
  url          = {https://www.nordicsemi.com/Products/nRF52840}
}

@Article{Ramirez2021BLE,
AUTHOR = {Ramirez, Ramiro and Huang, Chien-Yi and Liao, Che-An and Lin, Po-Ting and Lin, Hsin-Wei and Liang, Shu-Hao},
TITLE = {A Practice of {BLE RSSI} Measurement for Indoor Positioning},
JOURNAL = {Sensors},
VOLUME = {21},
YEAR = {2021},
NUMBER = {15},
ARTICLE-NUMBER = {5181},
URL1 = {https://www.mdpi.com/1424-8220/21/15/5181},
PubMedID = {34372415},
ISSN = {1424-8220},
DOI = {10.3390/s21155181}
}

\appendix

\subsection{Bernoulli Particle Filter Algorithm}

The \gls{pf} implementation of the \gls{bf} approximates the posterior spatial density using a set of weighted particles~$\{x^{(i)}_k, w^{(i)}_k\}_{i=1}^{N}$ and maintains the existence probability~$q_{k\mid k}$. 
\Algref{alg:bernoulli_pf} presents the complete implementation of the \gls{bf}.

\begin{algorithm*}[h]
\caption{A \gls{pf} Implementation of the \gls{bf} for Multi-Sensor Target Tracking with Missed Detections and False Alarms}
\label{alg:bernoulli_pf}
\begin{algorithmic}[1]

\State \textbf{Input:} Particles $\{x^{(i)}_{k-1}, w^{(i)}_{k-1}\}_{i=1}^{N+B}$, existence probability $q_{k-1\mid k-1}$, measurements $\mathcal{Z}_k = \big( Z^{(1)}_k, \ldots, Z^{(S)}_k \big)$.

\Statex \textbf{Existence Prediction:}
\State $q_{k\mid k-1} = p_b \cdot (1-q_{k-1\mid k-1}) + p_s q_{k-1\mid k-1}$

\Statex \textbf{State Prediction: (Survival Particles)}
\For{$i=1$ to $N$}
    \State Sample $x^{(i)}_k \sim f(x_k \mid x^{(i)}_{k-1})$ 
\EndFor

\Statex \textbf{State Prediction: (Birth Particles)}
\For{$i=N+1$ to $N+B$}
    \State Sample $x^{(i)}_k \sim b_k(x_k, \mathcal{Z}_k)$ using \Algref{alg:birth_init}
\EndFor

\Statex \textbf{Weight Prediction (Survival Particles):}
\For{$i=1$ to $N$}
\vspace{5pt}
    \State $w^{(i)}_{k\mid k-1} = \dfrac{p_s q_{k-1\mid k-1}}{q_{k\mid k-1}} w^{(i)}_{k-1}$
\EndFor

\Statex \textbf{Weight Prediction (Birth Particles):}
\For{$i=N+1$ to $N+B$}
    \State $w^{(i)}_{k\mid k-1} = \dfrac{p_b \cdot (1-q_{k-1\mid k-1})}{q_{k\mid k-1}} \dfrac{1}{B}$ 
\EndFor 

\Statex \textbf{Measurement Update:}
\For{$s=1$ to $S$}
    \State $I_1 = \sum_i P_{D,s}(x^{(i)}_k; \gamma) \, w^{(i)}_{k\mid k-1}$
    \For{each $z \in Z^{(s)}_k$}
        \State \!\!\!\!\!\! $I_2(z) \!= \sum_i P_{D,s}(x^{(i)}_k; \gamma) \, \ell_k^{(s)}(z \mid x^{(i)}_k, \gamma) \, w^{(i)}_{k\mid k-1}$
    \EndFor
    \State $\Delta_k^{(s)} = I_1 - \sum_{z \in Z^{(s)}_k} \dfrac{I_2(z)}{\kappa(z)}$
    
    \State $q_{k\mid k} = \dfrac{(1-\Delta_k^{(s)}) q_{k\mid k-1}}{1-\Delta_k^{(s)} q_{k\mid k-1}}$
    \vspace{2pt}
    \For{$i=1$ to $N+B$}
    \vspace{2pt}
        \State $w^{(i)}_{k\mid k} = w^{(i)}_{k\mid k-1} \Bigg[ 1 - P_{D,s}(x^{(i)}_k; \gamma) + \sum_{z \in Z^{(s)}_k} \dfrac{P_{D,s}(x^{(i)}_k; \gamma) \, \ell_k^{(s)}(z \mid x^{(i)}_k, \gamma)}{\kappa(z)} \Bigg]$
    \EndFor
    
    \State $w^{(i)}_{k\mid k} \leftarrow \dfrac{w^{(i)}_{k\mid k}}{\sum_j w^{(j)}_{k\mid k}}$ for $i=1$ to $N+B$
    \State $q_{k\mid k-1} \leftarrow q_{k\mid k}$
    \State $w^{(i)}_{k\mid k-1} \leftarrow w^{(i)}_{k\mid k}$ for $i=1$ to $N+B$
\EndFor

\Statex \textbf{Resampling:}
\For{$i=1$ to $N$}
    \State Draw $j \sim w_{k\mid k}$
    \State $x^{(i)}_k \leftarrow x^{(j)}_k$, 
    \State $w^{(i)}_k \leftarrow 1/N$
\EndFor

% \Statex \textbf{Birth Particles:}
% \For{$i=N+1$ to $N+B$}
%     \State Sample $x^{(i)}_k \sim p_b(x_k, Z_k)$
%     \State $w^{(i)}_k \leftarrow 1/B$
% \EndFor

\Statex \textbf{State Estimation:}
\If{$q_{k\mid k} > q_{\text{threshold}}$}
    \State $\hat{x}_k = \sum_{i=1}^N w^{(i)}_k x^{(i)}_k$
\Else
    \State No target detected
\EndIf

\end{algorithmic}
\end{algorithm*}

\subsection{Online Estimation of the Unknown Detection Probability Baseline}\label{sec:unknown_pd_appendix}
Following the unknown-detection-probability estimator of \cite{li2021robustpmbm}, adapted here to the single-target \gls{bf}, each sensor $s$ is assigned an independent belief over its detection probability, $P_{D,s} \sim \mathrm{Beta}\big(\alpha_k^{(s)},\beta_k^{(s)}\big)$, initialized with the uninformative prior $\alpha_0^{(s)}=\beta_0^{(s)}=1$ (mean $0.5$). 
The mean and variance of the Beta distribution are given by
\begin{subequations}
\begin{align}
    \mu_k^{(s)} &= \frac{\alpha_k^{(s)}}{\alpha_k^{(s)}+\beta_k^{(s)}}, \\
    v_k^{(s)} &= \frac{\alpha_k^{(s)}\beta_k^{(s)}}{\big(\alpha_k^{(s)}+\beta_k^{(s)}\big)^2\big(\alpha_k^{(s)}+\beta_k^{(s)}+1\big)}.
\end{align}
\end{subequations}

The belief is predicted and updated in the prediction and update steps of the particle filter described above, and its posterior mean
\begin{equation}\label{eq:pd_hat}
\hat P_{D,s} = \mu_k^{(s)} = \frac{\alpha_k^{(s)}}{\alpha_k^{(s)}+\beta_k^{(s)}}
\end{equation}
replaces the state-dependent $P_{D,s}(x;\gamma)$ everywhere it appears in the measurement update for sensor $s$, \eg, in \eqref{eq:req_update_full}.

\emph{Predict:} at every prediction step, the belief's mean is preserved but its variance is inflated by a factor $k_\beta \geq 1$, so that it keeps adapting to a possibly time-varying $P_D$ rather than collapsing to a point estimate:
\begin{subequations}
\begin{align}
\mu_{k \mid k-1}^{(s)} &= \mu_{k-1}^{(s)}, \\
v_{k\mid k-1}^{(s)} &= k_\beta \, v_{k-1}^{(s)}, \\
\alpha_{k\mid k-1}^{(s)} &= \nu_{k\mid k-1}^{(s)}\, \mu_{k-1}^{(s)}, \\
\beta_{k\mid k-1}^{(s)} &= \nu_{k\mid k-1}^{(s)}\, \big(1-\mu_{k-1}^{(s)}\big),
\end{align}
\end{subequations}
where
\begin{equation*}
    \nu_{k\mid k-1}^{(s)} = \frac{\mu_{k-1}^{(s)}\big(1-\mu_{k-1}^{(s)}\big)}{v_{k\mid k-1}^{(s)}} - 1, 
\end{equation*}
is the Beta distribution's concentration parameter, and $\mu_{k-1}^{(s)}$ and $v_{k-1}^{(s)}$ are the mean and variance of the Beta belief at the previous time step, and $\alpha_{k\mid k-1}^{(s)}$ and $\beta_{k\mid k-1}^{(s)}$ are the predicted parameters of the Beta distribution for sensor $s$ at the current time step.

\emph{Update:} let $\hat P_{D,s} = \mu_{k-1}^{(s)}$ (unchanged by the mean-preserving predict step above) be used in place of $P_{D,s}(x_k^{(i)};\gamma)$ in the per-particle bracketed term of \eqref{eq:req_density_update}, while the likelihood retains its threshold-truncated form $\ell^{(s)}(z\mid x_k^{(i)},\gamma)$, this yields the per-particle update weight
\begin{equation}
\Lambda_k^{(s),i} = 1 - \hat P_{D,s} + \underbrace{\hat P_{D,s} \sum_{z \in Z_k^{(s)}} \frac{\ell^{(s)}(z \mid x_k^{(i)}, \gamma)}{\kappa(z)}}_{D_k^{(s),i}},
\end{equation}
where $D_k^{(s),i}$ denotes only the detected term.
The particle-weighted ratio of these two quantities gives a soft credit assignment $\rho_k^{(s)} \in [0,1]$, the fraction of this step's evidence for sensor $s$ attributable to a detection rather than a missed detection,
\begin{equation}
\rho_k^{(s)} = \frac{\sum_{i} w_{k\mid k-1}^{(i)}\, D_k^{(s),i}}{\sum_{i} w_{k\mid k-1}^{(i)}\, \Lambda_k^{(s),i}},
\end{equation}
which is used to update the Beta belief, scaled by the predicted existence probability $q_{k\mid k-1}$ so that a step where the target is unlikely to exist barely moves the belief:
\begin{subequations}
\begin{align}\label{eq:beta_update}
\alpha_k^{(s)} &= \alpha_{k\mid k-1}^{(s)} + q_{k\mid k-1}\, \rho_k^{(s)}, \\
\beta_k^{(s)} &= \beta_{k\mid k-1}^{(s)} + q_{k\mid k-1}\, \big(1-\rho_k^{(s)}\big).
\end{align}
\end{subequations}
The updated posterior mean \eqref{eq:pd_hat} (with $k-1$ replaced by $k$) is then carried forward as $\hat P_{D,s}$ for the next time step. The inflation factor $k_\beta$ is chosen by a grid search over $k_\beta \in \{1.00, 1.10, 1.25, 1.50, 1.75, 2.00, 2.50, 3.00\}$ to minimize the average \gls{gospa} metric, with the best-performing value found to be $k_\beta = 2.00$, and $k_\beta = 1.10$ is a close second.

\end{document}